\documentclass[prb,twocolumn,superscriptaddress,showpacs]{revtex4} \usepackage{epsfig,braket}

\begin{document}

\title{Full protection of superconducting qubit systems from coupling
errors}
\author{M.J.\,Storcz\footnote{Email:
storcz@theorie.physik.uni-muenchen.de}}
\affiliation{Physics Department, ASC, and CeNS,
Ludwig-Maximilians-Universit\"at,  Theresienstr.\ 37, 80333 M\"unchen,
Germany}
\author{J.\,Vala} \author{K.R.\,Brown} \author{J.\,Kempe\footnote{also
affiliated with   CNRS-LRI, UMR 8623, Universit\'e de Paris-Sud, 91405
Orsay, France}} 
\affiliation{Department of
Chemistry and Pitzer Center for Theoretical Chemistry, University of
California, Berkeley, California 94720}
\author{F.K.\,Wilhelm}
\affiliation{Physics Department, ASC, and CeNS,
Ludwig-Maximilians-Universit\"at,  Theresienstr.\ 37, 80333 M\"unchen,
Germany}
\author{K.B.\,Whaley} \affiliation{Department of
Chemistry and Pitzer Center for Theoretical Chemistry, University of
California, Berkeley, California 94720}

\begin{abstract}
Solid state qubits realized in superconducting circuits  are
potentially extremely scalable.  However, strong decoherence may be
transferred to the qubits by various elements of the circuits that
couple individual qubits, particularly when coupling is implemented
over long distances. We propose here an encoding that provides full
protection against errors originating from these coupling elements,
for a chain of superconducting qubits with a nearest neighbor anisotropic
XY-interaction.  
The encoding is also seen to
provide partial protection against errors deriving
from general electronic noise.
\end{abstract}


\maketitle  Superconducting flux qubits have been shown to possess
many of the necessary features of a quantum bit (qubit), including the
ability to prepare  superpositions of quantum states
\cite{caspar,stonybrook} and manipulate them coherently
\cite{nakamura}.  In  these systems, the dominating error source
appears to be decoherence due to flux noise  \cite{assp}. Present
designs for arrays of {\it multiple} flux qubits that are coupled through
their flux degree of freedom are easily implemented from an
experimental point of view \cite{twofluxqb}.  However, when scaling up to
large numbers of qubits, they
suffer from technical restrictions such as possible flux crosstalk and
a need for physically large coupling elements, which are expected to act as 
severe antennas for 
decoherence.  The possibility of avoiding errors by prior encoding
into decoherence free subspaces (DFS) that are defined by the physical
symmetries of the qubit interaction with the environment is
consequently very attractive.  Such encoding is also attractive for
superconducting charge qubits \cite{Pashkin,Yamamoto}, which are subject to similar decoherence
sources \cite{zhou}.

In this letter, we show how to develop such protection for qubits
coupled by the nearest neighbor XY-interaction that is encountered in
both flux and charge qubit designs \cite{Levitov,Jens}. We demonstrate
that for this coupling, a two-qubit
encoding into a DFS provides full protection against noise from
the coupling elements.  Moreover, all encoded single-qubit operations
are also protected from collective decoherence
deriving from the electromagnetic environment.  The protection is seen
to result from a  combination of symmetry in the coupling element and a
restricted environmental phase space of the multi-qubit system --- the DFS
{\em alone} would not be sufficient. The
analysis makes use of an exact unitary transformation of $1/f$ {\em
phase} noise in the coupling element (hence with a sub-Ohmic power spectrum)
into regular nearest-neighbor correlated {\em flux} noise on the qubits
that is characterized by a super-Ohmic power spectrum. To assess
the performance of the encoding we add to this coupling-derived noise
a single-qubit Ohmic noise source that represents the generic uncorrelated
environmental factors and analyze the fidelity of encoded quantum gate 
operations.

The Hamiltonian of a linear chain of XY coupled qubits reads
\begin{eqnarray} \label{Hamiltonian}
\mathbf{H}_{q} & = & \mathbf{H}_0 + \mathbf{H}_{\rm int} \nonumber \\
& = & \sum_i \left[\epsilon_i \hat \sigma_z^{(i)} +
\Delta_i \hat \sigma_x^{(i)}\right.  \nonumber \\ {} &&\left. +
K_{i,i+1} \left(\hat \sigma_x^{(i)} \hat \sigma_x^{(i+1)}+ \hat
\sigma_y^{(i)} \hat \sigma_y^{(i+1)} \right)\right]\textrm{,} \quad
\end{eqnarray}
where $\mathbf{H}_0 = \sum_i \left[\epsilon_i \hat \sigma_z^{(i)} +
\Delta_i \hat \sigma_x^{(i)}\right]$ is the uncoupled qubit Hamiltonian,
and  $K_{i,i+1}$ is the strength of the inter-qubit coupling, $\mathbf{H}_{\rm int}$.   
We
assume that it is possible to switch the coupling $K_{i,i+1}$ and the
flux bias $\epsilon_i(\Phi_{x,i})$ of each qubit separately. Such a
Hamiltonian can be realized using flux qubits with capacitive coupling
\cite{Levitov}.  The switch for this interaction can in principle be
implemented using PIN varactor diodes, micromechanical devices, or
small Josephson junctions \cite{averinbruder}.  Switching on the coupling
suppresses the tunnel amplitudes\cite{Levitov} $\Delta_i$.
The Hamiltonian of Eq.\ (\ref{Hamiltonian}) can also be
readily implemented in charge qubits, i.e., Cooper pair
boxes coupled by Josephson junctions \cite{Jens}, whose coupling
strength can be tuned through an external magnetic field.  In
both cases, the couplers are large objects and hence act as  efficient
antennas for charge and/or flux noise when the coupling is on.
When the coupling is switched off, this noise is confined
within the coupler and does not affect the qubits.

The decoherence sources relevant to Eq.~(\ref{Hamiltonian}) are background
charges. This can be represented as $1/f$ noise in the coupler as we explain below.
In addition general electromagnetic (e.m.) noise, both local flux or electronics noise,
couples to single qubits and, for long wavelength, also to multiple qubits. 
The e.m. noise is represented as usual by Ohmic noise which has both uncorrelated and
collective components.
The effect of these environmental
decoherence sources on Eq.~(\ref{Hamiltonian}) is represented by the usual 
(linear) coupling to a bath of oscillators 
$\mathbf{H}_b=\sum_i \big(a_i^\dagger a_i+1/2\big)$, characterized by a spectral density
$J(\omega)=\sum_i \left|\lambda_i\right|^2\delta(\omega-\omega_i)$, with the 
coupling strength characterized by a dimensionless
parameter\cite{shnirman} $\alpha$.

We first show how the coupling and local noise are described in this framework.
Background charge fluctuations $\delta q (t)$ arising in the capacitive 
coupling elements 
between qubits $i$ and $i+1$, induce 
geometric Aharonov-Casher \cite{frankepl} phases
$\delta\phi(t)\propto\delta q(t)$ when the qubit flux states tunnel between
eigenstates of $\hat \sigma_z$.  This results in a correlated two-qubit
error operator $\exp\left[i\delta\phi\left(\hat \sigma_z^{(i)}+\hat
\sigma_z^{(i+1)}\right)\right]$ acting on $\mathbf{H}_{{\rm q}}$.
The low-frequency limit of this phase noise in the coupling elements
can be approximated as a Gaussian $1/f$ noise process
deriving from coupling to a sub-Ohmic oscillator bath
with associated spectral density\cite{shnirman,frankepl} 
$J^{2qb}_\phi(\omega)=(\alpha_0/\epsilon_0) {\rm sign}(\omega) e^{-\omega/\omega_c}$.
Here and henceforth we set $\hbar,k_B=1$.
This leads to a classical power spectrum in the frequency domain
\begin{eqnarray}
S_\phi(\omega) & = & \frac{1}{2} \braket{\delta \phi (t) \delta \phi (0)
+\delta \phi (0) \delta \phi (t) }_\omega \nonumber \\
 & = & J^{2qb}_\phi(\omega)\coth(\omega/2T)\simeq (2T\alpha_0/\omega_c\omega)
\end{eqnarray}
for $\omega\ll T$, which characterizes the environmental phase space
of the correlated two-qubit errors due to capacitive coupling.
Uncorrelated single qubit errors deriving from local
electronic elements are represented here by bath coupling to
the flux states, i.e., $\hat\sigma_z$ errors. This
is typically 
represented by a bath having an Ohmic spectral density\cite{markusfrankpra},
$J^{1qb}_{\epsilon,\Omega}=\alpha_\Omega\omega \omega_c^2 / (\omega_c^2+
\omega^2)$, which thus characterizes the
environmental phase space of the uncorrelated single-qubit errors.  
We note that very recently, 
$\hat \sigma_x$ single-qubit errors (i.e., bit flip errors)
have also been identified \cite{Robertson}. 
The third source of errors, correlated errors deriving from 
long wavelength electromagnetic radiation, 
can be removed by encoding into a DFS as we show below, independent of the
the form of the spectral density associated with the source of such
collective decoherence.

We can formally introduce the noise due to background charges into
the total Hamiltonian $\mathbf{H}_{q} + \mathbf{H}_{b}$ by transforming the total
Hamiltonian with a unitary operator
$U_{qb}=\exp\left[i\delta\phi\left(\hat \sigma_z^{(i)}+\hat
\sigma_z^{(i+1)}\right)\right]$, resulting in
\begin{equation}
\mathbf{H}=\mathbf{H}' + \mathbf{H}_b = U_{qb}\mathbf{H}_q U_{qb}^{\dagger}+ \mathbf{H}_b\textrm{, }
\end{equation}
with associated spectral density $J^{2qb}_\phi(\omega)$. Thus, the error acts in the
interaction picture as a time-dependent unitary transformation and it can be
eliminated by undoing the transformation. In NMR (nuclear magnetic resonance) language,
this is a transformation to the ``co-fluctuating'' frame.
The unitary transformation is properly undone by a time-dependent unitary
transformation in the interaction picture, which transforms the states
as $|\psi^\prime\rangle=U_{qb}^\dagger|\psi\rangle$ and the coupled
Hamiltonian as 
\begin{eqnarray}
\mathbf{H}_{\rm eff}&=&U_{qb}^\dagger \mathbf{H}
U_{qb}-i U_{qb}^\dagger \frac{d}{dt} U_{qb}, \\
U_{qb}^\dagger \frac{d}{dt} U_{qb} &=&\frac12 \left[{\hat \sigma}_z^{(i)}+{\hat \sigma}_z^{(i+1)}\right]\delta \dot \phi \textrm{.}
\end{eqnarray}
The last term is understood as  an effective system-bath 
interaction, written more explicitly
\begin{eqnarray} \label{HSB}
\mathbf{H}_{\rm SB} & = & U_{qb}^\dagger \frac{d}{dt} U_{qb} \nonumber\\
& = &\frac12 \left[{\hat \sigma}_z^{(i)}+{\hat \sigma}_z^{(i+1)}\right] \otimes \sum_n
i\omega_n\lambda_n\left(a_n-a_n^\dagger\right)\textrm{.}
\end{eqnarray}
Note that $\mathbf{H}_{\rm q} = U^\dagger_{qb} \mathbf{H}' U_{qb}$. Physically, this
arises from the transformation into the non-inertial co-fluctuating frame as an
inertial force. It is recognized that (\ref{HSB})
is the regular spin boson coupling $\mathbf{H}_{\rm SB,eff}=\sum_i (\lambda_i^\prime
a_i+\lambda_i'^{\ast}a_i^\dagger)$ with $\lambda_i^\prime=i\omega\lambda_i$.
In this transformed representation we now have correlated flux errors, i.e.,
pairwise coupling of the qubit $\hat \sigma_z$ operators to
energy fluctuations given by the 
time-derivative of the fluctuating correlated coupler phase, 
$\delta\dot{\phi}$.  Most importantly, the associated spectral density of the
oscillator bath is also transformed, becoming
$J^{2qb}_\epsilon(\omega)=\omega^2J^{2qb}_\phi(\omega)=\alpha_0 \omega^2{\rm
sign}(\omega)/\epsilon_0$, which is now super-Ohmic.  
Similar arguments can be applied to the flux noise
arising when two charge qubits are coupled by a SQUID, except 
that here the coupling (flux) noise is usually Ohmic rather than sub-Ohmic,
so that the transformed spectral density is proportional to $\omega^3$ 
rather than to $\omega^2$. Note, that the flux states 
only get transformed by phase factors, hence computation and measurement 
carried out in this basis are unaffected by this transformation.

To protect against these correlated errors 
we employ a two-qubit encoding $\ket{0}_L=\ket{01}$,
$\ket{1}_L=\ket{10}$  which is recognizable as the smallest DFS
encoding that can protect against collective dephasing \cite{kempe01}.
It therefore automatically protects against any correlated phase errors, 
including our third source of error deriving from long wavelength e.m.
noise.
We will show that this encoding also provides complete protection 
against the capacitive coupling noise, resulting in perfect 
performance of both 
encoded single qubit and two qubit operations when correlated errors
during two-qubit operations
are the only source of decoherence.  Uncorrelated
single qubit errors are then the only remaining mechanism leading to
a reduced fidelity of quantum gates.  We see below that for single qubit 
errors of
less than or equal strength to two qubit errors, the DFS encoding 
still provides a significant, although now incomplete, protection. 

The two logical qubits are encoded into four physical qubits using the
encoding scheme $\ket{00}_L=\ket{0101}_P$, $\ket{01}_L=\ket{0110}_P$,
$\ket{10}_L=\ket{1001}_P$, $\ket{11}_L=\ket{1010}_P$, where $L$ and $P$
denote logical and physical states, respectively.  We assume that the
four physical qubits constitute a linear array (this need not be
contiguous) which we label $1, 2, 3, 4$.  This four-dimensional
subspace is left invariant by collective errors involving qubits 1 and
2, $\hat \sigma_z^{(1)}+ \hat \sigma_z^{(2)}$, as well as by errors
involving qubits 3 and 4, $\hat \sigma_z^{(3)}+ \hat \sigma_z^{(4)}$,
but not by collective errors involving qubits 2 and 3, i.e., $\hat
\sigma_z^{(2)}+\hat \sigma_z^{(3)}$, see Ref. [\onlinecite{kempe01}].   The 
latter errors arise, when switching on the coupling between qubits 2 and 3 with
$\mathbf{H}_{{\rm int}}$ as described above, in order to perform  logical two-qubit operations.

The encoded single-qubit operations, given here without loss of
generality  for the first encoded logical qubit only, can be shown to be
\begin{eqnarray}
e^{-i\overline{\sigma}_z^{(1)} \tau} & = & e^{-i\hat
\sigma_z^{(2)} \tau}\\
e^{-i\overline{\sigma}_x^{(1)} \tau} & = & e^{-i\widetilde \mathbf{H}_{\rm
int}^{12}\tau}\\ 
e^{-i\overline{\sigma}_y^{(1)} \tau} & = &
e^{i\overline{\sigma}_z^{(1)} \frac{\pi}{4}}
e^{i\overline{\sigma}_x^{(1)} \tau}  e^{-i\overline{\sigma}_z^{(1)}
\frac{\pi}{4}},
\end{eqnarray}
where $\widetilde \mathbf{H}_{\rm int} = (\mathbf{H}_{\rm int}/\epsilon_0)$ and
$\tau=t \epsilon_0$. 
The first operation is straightforwardly achieved by tuning the flux bias.
To implement the second operation, $\overline{\sigma}_x^{(1)}$,
we need to cancel the effect of
$\mathbf{H}_0$.  This is also straightforward, if $\epsilon_i$ and $\Delta_i$ can be 
tuned to
zero.  If $\Delta$ can not be tuned to zero, it is nevertheless
still possible to act with $\mathbf{H}_{\rm int}$ alone, by combining a short time 
Trotter expansion with operator conjugation as follows. 
First, we recognize that 
conjugation of $\mathbf{H}_q$ with $\hat \sigma_z$ can invert
the sign of $\Delta_i$
\begin{eqnarray}
e^{-i\mathbf{H}_q(-\Delta_1,-\Delta_2) t} & = & e^{-i(\hat
\sigma_z^{(1)}+\hat \sigma_z^{(2)}) \pi /2} \times {} \nonumber \\
&& {} \times 
e^{-i\mathbf{H}_q(\Delta_1, \Delta_2)t} e^{i(\hat \sigma_z^{(1)}+ \hat
\sigma_z^{(2)}) \pi /2}\textrm{.}\qquad
\end{eqnarray}
The alternation of $\mathbf{H}_q(\Delta_1,\Delta_2,K_{12})$ with 
$\mathbf{H}_q(-\Delta_1,-\Delta_2,K_{12})$ results in the desired
action of $\mathbf{H}_{\rm int}$, up to 
commutator errors between $\mathbf{H}_{\rm int}$ and $\Delta \hat \sigma_{x}^{(i),(j)}$ 
which can be suppressed by making a Trotter expansion:
\begin{eqnarray} \label{bb}
& & \lim_{n \rightarrow \infty} \Big( e^{-i\mathbf{H}_q(\Delta_1, \Delta_2,
K_{12}) t  /2n } e^{-i\mathbf{H}_q(-\Delta_1,-\Delta_2,
K_{12}) t /2n } \Big)^n {} \nonumber\\
&& {} =  e^{-i\widetilde \mathbf{H}_{\rm
int}^{12}\tau}\textrm{.}
\end{eqnarray}
This scheme requires only relatively small values of $n$ to be
effective. Direct simulation shows that for $n \sim 10$, 
the relative deviation of individual
matrix elements $U_{nm}$ from $U_{nm}^{\rm ideal}$ is smaller than 1\%.
During all these encoded single qubit operations the encoded qubit remains in 
the DFS encoded
subspace and so is fully
protected against correlated two-qubit errors deriving from
both the capacitive coupling and from any other electromagnetic  
correlated noise.
\begin{figure}[t]
\begin{center}
 \includegraphics*[width=8cm]{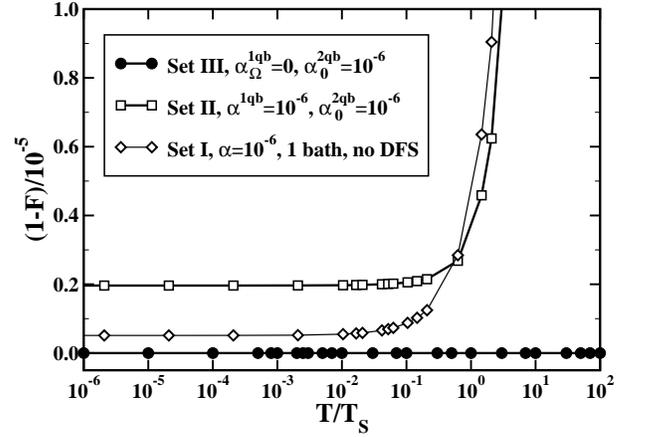}
\end{center}
        \caption{Fidelity deviation $1-{\mathcal F}$ versus temperature for the encoded 
          CNOT operation, shown for two 
        different combinations of super-Ohmic two-qubit noise (strength 
        $\alpha^{\rm 2qb}=\alpha_0$) and
        Ohmic single-qubit noise (strength 
        $\alpha^{\rm 1qb}=\alpha_\Omega$) acting on the physical qubits.  
        The characteristic
        temperature scale is $T_s=\epsilon_0(h/k_B)$, yielding
        $T_s=48$ mK for qubits with energies
        $\epsilon_i = \epsilon_0 \equiv 1$~GHz, $i=1,2$. Here, $\epsilon_0$ 
 is used as an energy unit for the correlation function.
Solid lines are provided as guides to the eye.
        Ideal gate performance is achieved when $\alpha^{\rm 1qb}=0$. 
        Detailed analysis shows that the
        fidelity depends linearly on $\alpha^{\rm 1qb}$.
        For comparison, set I shows the corresponding performance of 
        the unencoded CNOT operation taken from Ref.~\onlinecite{markusfrankpra}.}
        \label{CNOTSuperohmic}
\end{figure}

Encoded two-qubit operations require pairwise coupling of physical
qubits from the two encoded qubits $\ket{0}_L$ and $\ket{1}_L$, e.g.,
qubits 2 and 3 as mentioned above. The
encoded $\overline{U}_{zz}(t)$ two-qubit controlled-phase operation is
\begin{eqnarray} \label{zz}
\overline{U}_{zz}(t)  & = &
 e^{-i\overline{\sigma}_z^{(1)}\overline{\sigma}_z^{(2)}\tau} \nonumber
 \\ & = &  e^{iS_x\frac{\pi}{4}}e^{-i\widetilde \mathbf{H}_{\rm
 int}^{23}\tau/2}e^{-i\sigma_x^{(2)}\frac{\pi}{2}} e^{i \widetilde \mathbf{H}_{\rm
 int}^{23}\tau/2} e^{iS'_x\frac{\pi}{4}} \textrm{, }
\end{eqnarray}
where $S'_x=\hat \sigma_x^{(2)}-\hat \sigma_x^{(3)}$ and $S_x=\hat
\sigma_x^{(2)}+\hat \sigma_x^{(3)}$.  This can be combined with an encoded single
qubit Hadamard gate to produce the controlled NOT (CNOT) gate \cite{markusfrankpra}.
Now the
first element of $\overline{U}_{zz}(t)$, $e^{iS'_x\frac{\pi}{4}}$, 
takes the DFS states outside the subspace to form superpositions of
DFS and non-DFS states and populate the non-DFS
states $\ket{0111}$, $\ket{0100}$, $\ket{1011}$ and
$\ket{1000}$. Detailed analysis reveals that the two-qubit operation
eq.\ (\ref{zz}) will always take the encoded qubits out
of the DFS encoded subspace. However, during these excursions out of the DFS, 
when only coupling errors are present, only pure
dephasing processes which do not flip eigenstates
can contribute to decoherence \cite{markusfrankpra}, since the coupling to
the bath commutes with the interqubit coupling.
The rates of these dephasing processes
are proportional to $S(0)=\lim_{\omega\rightarrow0}
J_\epsilon^{2qb}(\omega)\coth(\omega/2T)$, which vanishes as a result of the
super-Ohmic shape of $J^{2qb}_\epsilon$ derived from the tunneling-flux
transformation introduced above.  Consequently these processes 
''lack phase space''
in the environmental degrees of freedom and hence are fully
suppressed. This excursion out of the DFS encoded subspace into a
larger region of the full Hilbert space in which only pure dephasing
processes contribute to the decoherence can alternatively be viewed as
an excursion into a larger subspace that is
characterized by suppression of relaxation processes.

We demonstrate the benefits of the DFS encoding by numerical studies 
of the CNOT gate,
calculated from the simulated evolution of the reduced density matrix
for the coupled flux qubits using the Bloch-Redfield description of the
spin-boson  model of the qubit and its bath coupling characterized by
\cite{markusfrankpra} $J(\omega)$. This approach is valid for 
$\alpha_0,\alpha_\Omega\ll1$. 
To quantify the gate performance we
evaluate the fidelity \cite{GateQualityFactors} $\mathcal{F}$ of the
encoded quantum gate operation, defined by $ \mathcal{F} =
\frac{1}{16} \sum_{j=1}^{16}
\braket{\Psi_{in}^j|U_{G}^+\rho^j_{G}U_{G}|\Psi^j_{in}}$.  Here $U_G$
is the unitary matrix describing the desired ideal gate, and
$\rho^j_{G}=\rho(t_{G})$  is the density matrix obtained from
attempting a quantum gate operation in a hostile environment, 
i.e., with errors,
evaluated for all initially unentangled product states \cite{GateQualityFactors} from the
encoded logical basis, $\rho(0)=\ket{\Psi_{in}^j}\bra{\Psi_{in}^j}$. 
The states $\ket{\Psi_{in}^j}$ are defined in Ref.\,[\onlinecite{thorwart}].

Figure \ref{CNOTSuperohmic} shows the calculated gate fidelity
for an encoded CNOT operation $\overline{U}_{\rm CNOT}$, obtained from
$\overline{U}_{zz}$ together with the relevant encoded single qubit gates.
We see that, as predicted by the above analysis, when only two-qubit errors
are active ($\alpha^{\rm 1qb}=0$) the gate performance is perfect.  When additional
uncorrelated single-qubit errors during single qubit operations occur
($\alpha^{\rm 1qb}$), the gate fidelity is seen to decrease as the strength
of these errors increases. The DFS encoding is thus seen to give 100\% protection
against the primary coupling errors in addition to correlated background errors.  
It does not protect against  
uncorrelated single qubit errors, in fact, 
due to the larger overhead, DFS encoding alone is sensitive
against these (compare set I and set II).
However, the uncorrelated single qubit errors can be well treated 
by active quantum error correction, particularly if the error rates for single qubit
and correlated errors are comparable. It is also possible to combine this encoding scheme with a QECC in order
to achieve fault-tolerance. Using the scheme proposed in Ref.\,[\onlinecite{mohseni}],
the leakage problem of standard QECC methods can be overcome.

Saturation of the gate quality at low temperatures occurs because
all decohering processes (except spontaneous emission) are frozen out. This 
occurs when $k_BT\simeq E_{\rm min}$, where $E_{\rm min}$ is the lowest
energy splitting in the system. Here, $E_{\rm min}=\epsilon_0$.
Even during the excursion out of the DFS, transitions 
between the eigenstates of the Hamiltonian involving spontaneous emission 
are forbidden by symmetry. 
Thus, at low temperatures, only energy-conserving ``pure dephasing'' processes
influence the gate. These are proportional to the noise power $S(\omega\rightarrow0)$.
For an Ohmic environment, this noise is purely thermal\cite{markusfrankpra}, $S(0)\propto T$,
so that the gate performance is still limited at any finite temperature. For
the super-Ohmic case, $S(0)=0$ at any $T$ (figure 1).
When $\alpha_\Omega$ is small, the fidelity can be considerably increased because
the errors from the coupling elements introduce no new constraints; i.e.,
if, for equal coupling strength to the electromagnetic environment,
the appropriate relative weight of two qubit errors is larger than that of one 
qubit 
errors, it is evident that the DFS encoding provides considerable protection.
Thus, for optimizing two-qubit gates it is of crucial importance to identify,
whether or not the noise is correlated between qubits. 
This is a critical challenge for experiment.
An experimental signature of correlated noise is, e.g., the superior coherence of the 
states used as logical qubits in this letter. 

In conclusion, we have shown that using a DFS encoding of
superconducting flux or charge qubits can significantly enhance their
gate performance for the entangling  two-qubit operations that are
required to implement quantum computation. The DFS-encoding proposed
here ensures that  all {\em encoded} single-qubit operations are
protected against $1/f$ noise in the capacitive coupling elements,
as well as from correlated
electromagnetic noise. The latter are the errors originating from the
coupling of the qubits to a common electromagnetic environment.
When only the capacitive coupling errors arising during two-qubit operations are present, 
perfect fidelity can be achieved. We have shown that this results from 
an exact correspondence of the $1/f$ sub-Ohmic phase noise in the
coupler to super-Ohmic flux noise on the qubits.
Encoding is seen to offer a significant improvement
of the gate quality due to the supression of spontaneous emission. 
The phase space
restriction seen here derives from the choice of the XY-interaction between
the qubits: coupler noise from other interactions, such as ZZ, would
explore the full phase space during the two-qubit operation. Thus the XY-coupling is a
very attractive coupling scheme whenever decoherence is a major concern.

We expect that this DFS-inspired encoding,
which is also very efficient, requiring only two physical qubits per
logical qubit, will therefore be useful for reducing the noise in
quantum circuits based on superconducting qubits.

We thank T.L.\,Robertson for useful discussions.
This work was supported in part by the NSA and ARDA under ARO contract
number P-43385-PH-QC, in part by the National Science Foundation
under the ITR program, grant number EIA-0205641, and by DFG through
SFB 631.

\end{document}